\documentclass[12pt]{iopart}

\usepackage{graphicx}  
\usepackage{dcolumn}   
\usepackage{bm}        
\usepackage{bbm}       
\usepackage{iopams}
\usepackage{multirow}
\expandafter\let\csname equation*\endcsname\relax
\expandafter\let\csname endequation*\endcsname\relax
\usepackage{amsmath}
\newcommand{\erdoes}{Erd\"os-R\'{e}nyi }
\newcommand{\erdoesnospc}{Erd\"os-R\'{e}nyi}
\newcommand{\barabasi}{Barab\'{a}si-Albert }

\begin{document}
\title{Unimodular lattice triangulations as small-world and scale-free random graphs}
\author{B Kr\"{u}ger, E M Schmidt and K Mecke}
\address{Institute for Theoretical Physics, Universit\"at Erlangen-N\"urnberg, Staudtstr. 7, 91058 Erlangen}
\ead{benedikt.krueger@fau.de}

\begin{abstract}
  Real-world networks, e.\,g.\,the social relations or world-wide-web graphs, exhibit both small-world and scale-free behaviour.
  We interpret lattice triangulations as planar graphs by identifying triangulation vertices with graph nodes and one-dimensional simplices with edges.
  Since these triangulations are ergodic with respect to a certain Pachner flip, applying different Monte-Carlo simulations enables us to calculate average properties of random triangulations, as well as canonical ensemble averages using an energy functional that is approximately the variance of the degree distribution.
  All considered triangulations have clustering coefficients comparable with real world graphs, for the canonical ensemble there are inverse temperatures with small shortest path length independent of system size.
  Tuning the inverse temperature to a quasi-critical value leads to an indication of scale-free behaviour for degrees $k \geq 5$.
  Using triangulations as a random graph model can improve the understanding of real-world networks, especially if the actual distance of the embedded nodes becomes important.
\end{abstract}
\submitto{\NJP}

\maketitle

\section{Introduction}
Real world systems or networks that consist of many similar entities that are interacting can be described by graph theory (see \cite{Newman_2010,Costa_2011} for reviews).
Examples are the world-wide-web, where single websites are modelled as vertices and links between websites are modelled as edges;
the network of social interactions, where vertices represent single humans and edges imply an existing friendship between two humans 
and the scientific co-author network, where authors of scientific papers are vertices and an edge between two authors is existent if both are co-authors of a common paper.

Most of these real networks share three common properties \cite{Albert_2002,Costa_2011}:
\begin{enumerate}
  \item The clustering coefficient, which is basically the probability that two neighbours of a common node are connected by a graph edge, is high and independent of the size of the network.
    In real networks this means for example that there is a high probability that if two persons know a common person, they know each other, too; if two scientific authors are co-authors of a common third author, also these two are very likely co-authors.
  \item The length of the shortest path between two random vertices, measured by the number of edges the path consists of, is small even for huge graphs and scales like the logarithm of the system size for increasing system size.
    This behaviour is commonly known for the network of social interactions, where every two random persons of the world should know each other over less then 10 middlemen, or in the world wide web, where each website can be reached from every other website within a few clicks.
  \item The distribution of the number of edges incident with a vertex (its degree $k$) follows a power-law distribution ($P(k) \propto k^{-\gamma}$) with $2 < \gamma < 3$, which means that there are a lot of vertices with only a few neighbours, and only a minority highly connected vertices.
    For example in the co-author network there are few authors that worked together with a lot of different people, and there are a lot of authors that worked only with very few co-authors.
\end{enumerate}
The appearance of the first two properties is often denoted as small-world behaviour, networks with the third property are denoted as scale-free.

To understand the structure and behaviour of real world networks different random graphs are used as model systems \cite{Albert_2002}.
Three widely used models are the \erdoes random graph \cite{Erdoes_1959,Erdoes_1960,Erdoes_1961}, the Watts-Strogatz random graph \cite{Watts_1998} (and a slightly altered version denoted as Newman-Watts random graph \cite{Newman_1999}) and the \barabasi random graph \cite{Barabasi_1999}.
While all three models have a shortest path length scaling with the logarithm of the system size \cite{Fronczak_2004,Watts_1998}, none exhibits both the other two properties of real-world networks:
The \erdoes random graph has vanishing clustering coefficient and a binomial degree distribution;
the Watts-Strogatz random graph exhibits a non-vanishing clustering coefficient for a special choice of parameters \cite{Watts_1998}, but also a degree distribution that agrees quantitatively with the \erdoes graph \cite{Barrat_2000};
the \barabasi model creates a scale-free degree distribution with exponent $\gamma = 3$ \cite{Barabasi_1999,Dorogovtsev_2000,Krapivsky_2000}, but leads also to a vanishing clustering coefficient for increasing system size \cite{Fronczak_2003}.
Several more complicated models developed afterwards combine both small-world and scale-free behaviour \cite{Holme_2002,Klemm_2002,Szabo_2003}, there are also models that use graphs with vertices embedded in some geometric space that exhibit these two properties \cite{Rozenfeld_2002,BenAvraham_2003,Xu_2007}.

We present here a novel type of (embedded) random graphs by identifying triangulations of integer lattice point sets as graphs that shows a crossover from ordered, large-world to unordered, small-world and possible scale free behaviour.
Such triangulations are tessellations of the convex hull of the point set into non-intersecting simplicial building blocks (triangles in two dimensions, tetrahedrons in three dimensions) \cite{DeLoera_2010}.
Triangulations are an important tool in physics for describing curved space(-times), in quantum geometry (e.g. in the framework of Causal Dynamical Triangulations \cite{Ambjorn_2005} and in spin foams \cite{Rovelli_2007}); they are also a major object of study in topology and geometry where one is for example interested in the number of distinct triangulations of a given topological manifold \cite{Sulanke_2009}.
Triangulations are commonly used for describing foams \cite{Sullivan_1999}, where often a special so-called Delaunay triangulation, which is the dual to the Voronoi tessellation, is used for foam construction.
They can also be used for describing the topological properties of foams in terms of the neighbouring cells \cite{Oguey_2003,Aste_1995,Dubertret_1998b}.
Additionally glass-like dynamics near regular configurations can be found in dual graphs of topological triangulations \cite{Aste_1999}.
The dual graphs of disordered triangulations as described in this paper can be used for the construction of poly-disperse foams, which exhibit a broad range of cell sizes, and for considering e.g. their transport and diffusion problems.

Triangulations have been used as random graph models before: 
Random Apollonian Networks \cite{Andrade_2005,Zhou_2005,Song_2012} (randomised dual graphs of Apollonian packed granular matter) are triangulations constructed using a procedure similar to preferential attachment and show both small-world and scale-free behaviour. 
Since each graph can be embedded into a closed surface with high enough genus, and triangulations are maximal planar graphs in the sense that each insertion of another edge will break planarity, triangulations of surfaces with arbitrary genus were studied in \cite{Aste_2012}.
Canonical ensembles of triangulations of the sphere were used in \cite{Kownacki_2004} to consider a quench from the random triangulations to zero temperature.

In contrast to the triangulations considered in \cite{Andrade_2005,Zhou_2005,Song_2012,Aste_2012,Kownacki_2004} before, which we denote as topological triangulations, we use embedded triangulations with vertices having fixed coordinates, precisely unimodular triangulations of two-dimensional integer point lattices.
For topological triangulations only the topological degrees of freedoms (the way how vertices are connected) are important, contrary to embedded triangulations where additionally the actual coordinates of the vertices are fixed and specified.
This causes complexities and numerical difficulties that do not have to be addressed in topological triangulations, e.g. to determine whether a Pachner flip leads to a valid triangulation.

In this paper we measure the degree distribution, the clustering coefficient and the shortest path length of random lattice triangulation graphs using Metropolis Monte-Carlo simulations and find a high clustering coefficient and similarities to common network models.
Introducing the notion of an energy of a triangulation that is well known in literature and corresponds to the variance of the degree distribution we apply the usual notion of statistical physics and examine the canonical ensemble averages of these observables for different values of the inverse temperature.
For the numerical calculation of the expectation values we use the Wang-Landau algorithm for calculating the density of states.
This makes it possible to calculate equilibrium properties for all temperatures, in contrast to Metropolis or Glauber dynamics used in \cite{Aste_1999,Aste_2012,Kownacki_2004}, where it is hard to access negative temperatures and low temperatures, but we loose the ability to consider dynamical behaviour like quenches or glass-transitions studied in the literature before.
In the canonical ensemble we find in all considered observables a transition from an ordered large-world behaviour for positive temperatures to small-world and scale free behaviour for negative temperatures.

\section{Triangulations as random graphs}\label{sec:triangulation_graphs}

\begin{figure}
  \begin{indented}
    \item[]{
        \includegraphics[width=4.0cm]{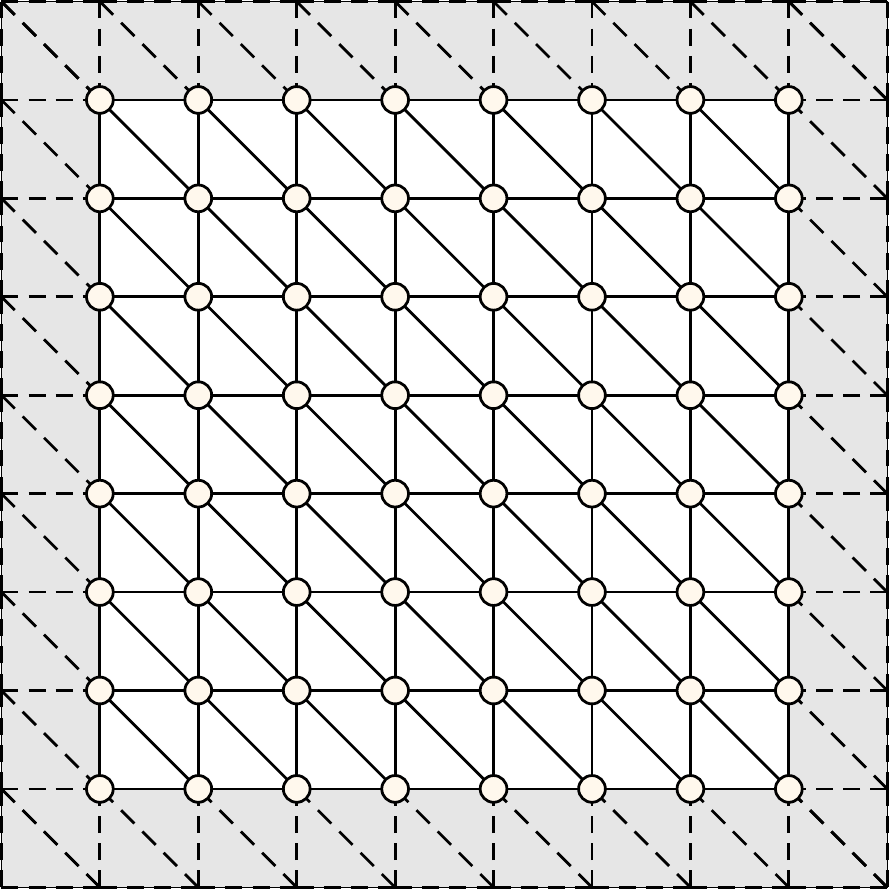}
        \hspace{0.4cm}
        \includegraphics[width=4.0cm]{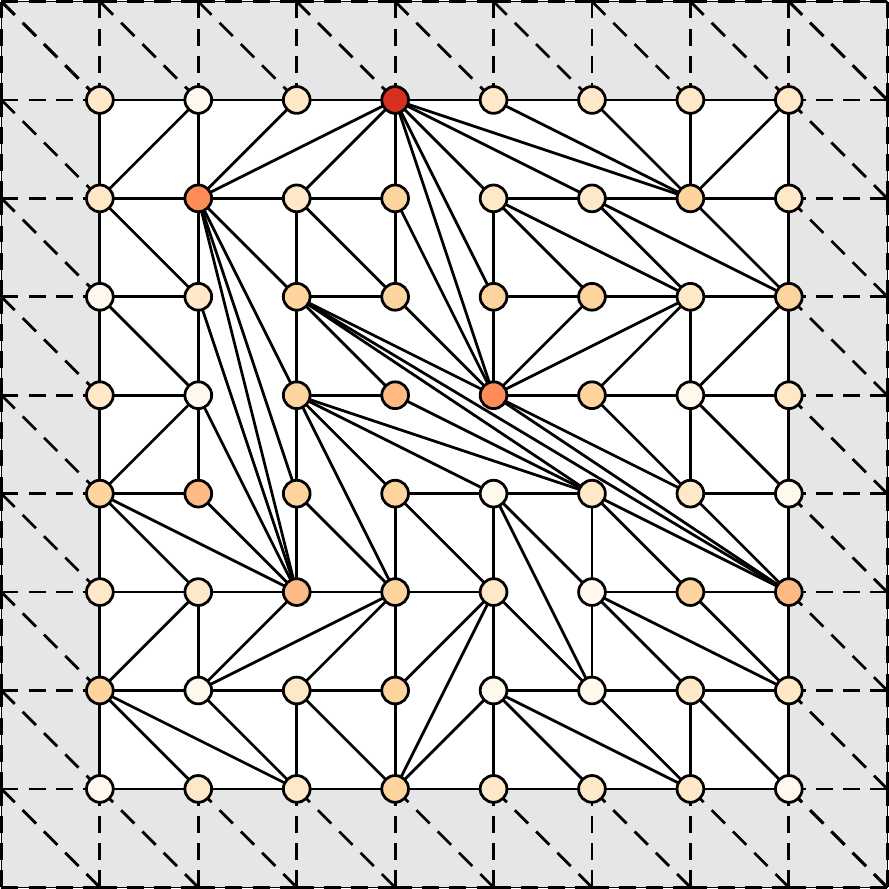}
        \hspace{0.4cm}
        \includegraphics[width=4.0cm]{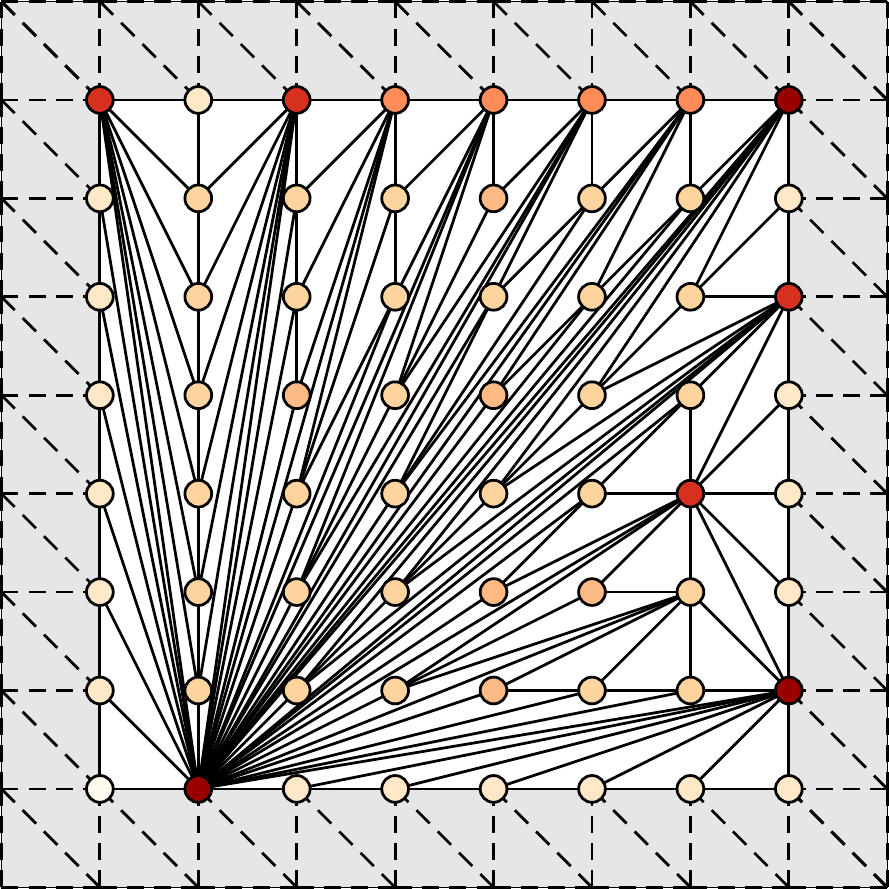}\\
        \makebox[4.0cm]{$\beta \rightarrow +\infty$}
        \hspace{0.4cm}
        \makebox[4.0cm]{$\beta = 0$}
        \hspace{0.4cm}
        \makebox[4.0cm]{$\beta \rightarrow -\infty$}
      }
  \end{indented}
  \caption{\label{fig:triang_examples}Examples of unimodular triangulations of a $8\times8$ lattice. The colour of the vertices corresponds to the number of incident vertices, the dashed border is fixed and will not be flipped, but is considered for the calculation of the energy. From left to right: Minimal energy triangulation ($\beta \rightarrow +\infty$), random triangulation at infinite temperature ($\beta = 0$), example of a maximal energy configuration ($\beta \rightarrow -\infty$). The colour code corresponds to the difference of the actual and the reference vertex degree as defined in (\ref{eq:energy}).}
\end{figure}

A triangulation of a discrete point set $\mathbf A = \{x_1, \dots x_K\}, x_i \in \mathbb R^2$ is a tessellation of the convex hull of the point set into triangles $\sigma_i \in \mathbf A^3$ so that the interiors of two distinct triangles $\sigma_i \neq \sigma_j$ do not intersect, and so that every point $X_i$ is the corner of at least one triangle \cite{DeLoera_2010}.
Sometimes the latter property is not required, if all points are corners of a triangle the triangulation is then called full, if there are points that are not part of the triangulation it is called non-full.
In this paper we consider only full triangulations of two-dimensional integer $M\times N$-lattices
\begin{equation*}
  \mathbf A = \left\{ \begin{pmatrix} m \\ n \end{pmatrix} \mid m \in Z_M := \{0, 1, \dots, M-1\}, n \in \mathbb Z_N \right\}
\end{equation*}
which are unimodular, that is that all triangles have equal area $1/2$. 
Examples of such triangulations can be found in figure\,\ref{fig:triang_examples}.

There are analytical bounds \cite{Kaibel_2003} and numerical calculations \cite{Knauf_2013} for the total number of triangulations on such integer lattices.
Both the bounds and the numerical calculations show that number of triangulations scales exponentially with the system size of the underlying lattice, this extensivity makes it possible to apply tools from statistical physics to triangulations.
A similar system was also considered in \cite{Caputo_2013}, where the convergence of a Metropolis-like Monte Carlo algorithm applied to lattice triangulations was analysed for different parameter choices.

Taking the edges and the corners of triangulations in addition to triangles into account as simplices, a triangulation is also a simplicial complex, i.e. all faces of simplices are also contained in the simplicial complex.
We interpret triangulations as graphs by identifying vertices as graph nodes and the triangulation edges as graph edges.
Neglecting the boundary vertices triangulation graphs are maximal planar, i.e. no edge between an internal and another vertex can be inserted without violating the planarity of the graph.

\begin{figure}
  \begin{indented}
  \item[]{\includegraphics[width=13.2cm]{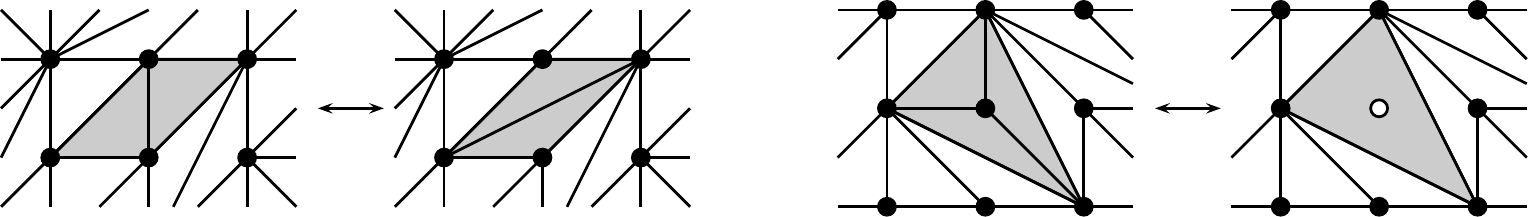}}
  \end{indented}
  \caption{\label{fig:pachner_moves}The two different Pachner moves in two-dimensional lattice triangulations. (Left) the diagonal edge flip that replaces one diagonal of a quadrilateral with the other diagonal. On the lattice this steps leads to a valid triangulation if the two triangles incident with the edge to be flipped constitute a parallelogram. (Right) the removal (up to down) and the insertion step, which are not used in our system, since we consider only full triangulations that are ergodic with respect to the diagonal edge flip.}
\end{figure}

Since we are not interested in properties of certain triangulations but in averages over all triangulations of a given point set, we need a method that is able to construct all possible triangulations of a given point set.
Therefore we use Pachner flips \cite{Pachner_1986}, elementary steps that map a triangulation into another one:
Select an edge $\{x_i, x_j\}$ of the triangulation that connects two neighbouring triangles $\{x_i, x_j, x_k\}$ and $\{x_i, x_j, x_l\}$.
If the quadrangle consisting of the two triangles is convex, replace the two triangles by $\{x_i, x_k, x_l\}$ and $\{x_j, x_k, x_l\}$; if the quadrangle has a concave or a flat angle, discard the flip as non-executable (see figure\,\ref{fig:pachner_moves} for some examples).
These flips are ergodic for triangulations of point sets in two dimensions \cite{Lawson_1972}, that is that every triangulation of a given point set can be transformed into every other one by a finite number of Pachner flips, so we can use these Pachner moves for doing Markov chain Monte Carlo simulations.
Notice that the introduced Pachner flip conserves the number of edges and the unimodularity of the triangulation.
In principle there are more Pachner moves (inserting and removing vertices) beside flips, but these lead to non-full triangulations and thus are not considered here.
In the context of graph theory on topological triangulations \cite{Kownacki_2004,Aste_2012} the Pachner moved here is called $T_1$, and the two additional moves not used here are called $T_2$.
Note that there is a fundamental difference between the diagonal edge flips in topological and in lattice triangulations: 
Each former flip can always been executed, whereas latter ones can only be executed if the two adjacent triangles form a convex quadrangle.
This is due to the fixation of the vertex coordinates in lattice triangulations, while for topological triangulations no coordinates of the vertices are specified and the notions of convex or concave do not exist.

To categorise triangulations of a point set according to their order and disorder, we define for a triangulation $\mathcal T$ the energy function
\begin{equation}\label{eq:energy}
  E(\mathcal T) := \sum_{v \in \mathbf A} \left( k_v^{(\mathcal T)} - k_v^{(\mathcal T_0)} \right)^2
\end{equation}
taking low values for ordered triangulations and high values for unordered triangulations.
Here $k_v^{(\mathcal T)}$ is the number of edges that are incident with the vertex $v$ in triangulation $\mathcal T$ and $\mathcal T_0$ is a reference triangulation that will be the ground-state of the energy function (in this paper we choose the maximal ordered triangulation displayed in figure\,\ref{fig:triang_examples} as the reference triangulation).
The additional term $k_v^{(\mathcal T_0)}$ with $\mathcal T_0$ being the maximal ordered triangulation can be seen as an implementation of fixed boundary conditions as displayed in figure\,\ref{fig:triang_examples}.
Periodic boundary conditions, which yield toroidal topology, cannot be used for lattice triangulations, because on closed surfaces without boundaries there are several possibilities to connect two vertices with an edge (e.g. with different winding number) and therewith it depends on the choice of the connection whether the considered object is a valid triangulation.
In contrast, for the considered topology, the edge between two vertices is always the line segment.

Since the number of edges of full lattice triangulations is constant, sums over linear vertex degrees vanish and the chosen energy function (\ref{eq:energy}) is the simplest possible polynomial energy function in the number of vertices, edges and triangles.
It can be related to the square of the local curvature usually used in dynamical triangulations \cite{Ambjorn_2005} (which is basically the deficit angle given by the number of triangles around a vertex minus six).
Similar energy functions were already applied to graphs in \cite{Farkas_2004}; 
in \cite{Caputo_2013} the total length of all edges was used as energy function in lattice triangulations, which qualitatively agrees with our choice since high energy leads to long edges in the triangulations;
previous works \cite{Aste_1999,Kownacki_2004,Aste_2012} considering topological triangulations as graphs also use the energy function (\ref{eq:energy}) with the mean vertex degree $\langle k \rangle$ instead of $k_v^{(\mathcal T_0)}$.

Interpreting the triangulation vertices as graph nodes and the 1-simplices as graph edges, we compare the graph properties of the triangulations with \erdoes \cite{Erdoes_1959,Erdoes_1960,Erdoes_1961}, Newman-Watts \cite{Newman_1999} and \barabasi \cite{Barabasi_1999} random graphs. 
In all three cases we choose the model parameters so that the number $n$ of the random graph nodes matches the number $MN$ of the lattice triangulation vertices and the average number of random graph edges matches the number $e = 3MN - 2(M+N) + 1$ of the lattice triangulation edges.

In the \erdoes random graph there are $n$ vertices, each pair of vertices is connected by an edge with probability $p$ \cite{Erdoes_1959,Erdoes_1960,Erdoes_1961}. 
They have a small world behaviour for the shortest path length, but vanishing clustering coefficient and a binomial degree distribution.
To compare the \erdoes random graph with triangulations we choose its parameters to be $n = MN$ and $p = 2e / n(n-1) \rightarrow 6/n$.

The Newman-Watts random graph \cite{Newman_1999} is a modification of the Watts-Strogatz graph \cite{Watts_1998}.
Starting with a regular graph of $n$ vertices and connections to the next $L$ neighbours, for each present edge an additional edge is inserted between two random vertices with probability $q$. 
This model can be seen as the periodic regular graph superimposed with a random graph.
The random rewiring leads to shortcuts and a short average path length, the basic regular graph leads to a high clustering coefficient.
To obtain a comparable random graph we use the parameters $n = MN$, $L=4$ and $q = -1 + e/2MN$

The \barabasi random graph \cite{Barabasi_1999} is constructed as following:
Start with $m$ isolated vertices and iteratively insert $t$ vertices, each with $m$ edges to already present vertices, such that the probability for connecting to a present vertex is proportional to its degree (preferential attachment).
This random graph has vanishing clustering coefficient, but a power-law degree distribution and a small-world shortest path length.
We choose the model parameters $m = 3$ and $t = MN - 3$ to compare with $M\times N$ lattice triangulations.

\section{Random triangulations}

\begin{figure}
\begin{indented}
\item[]{\includegraphics{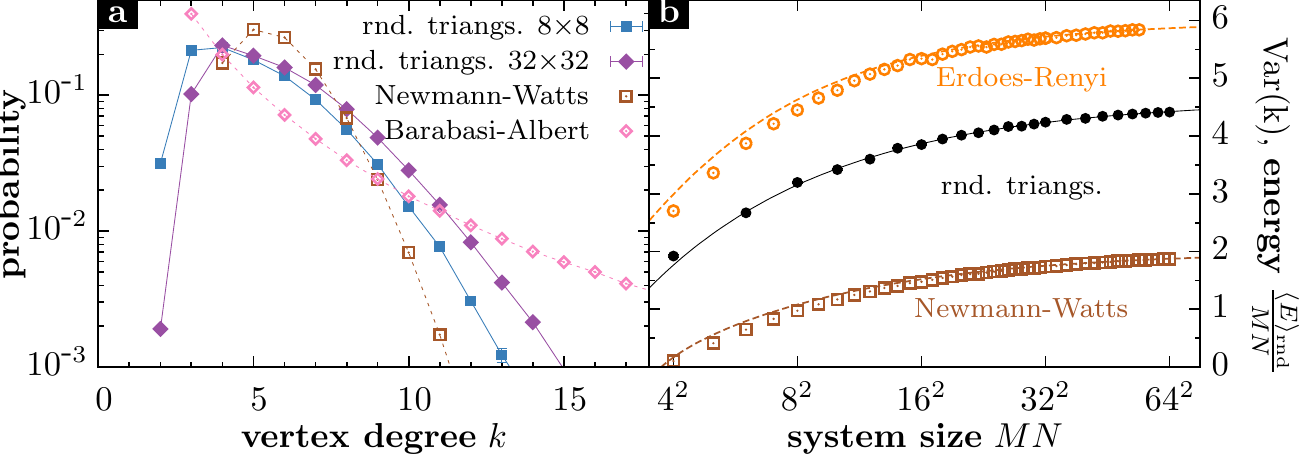}}
\item[]{\includegraphics{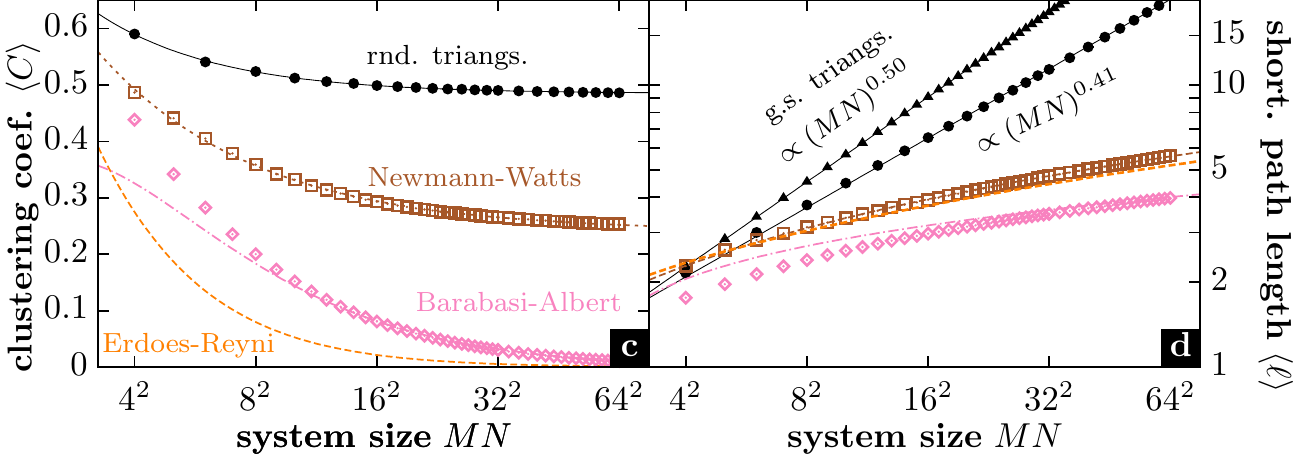}}
\end{indented}
\caption{Scaling behaviour of random lattice triangulations and comparable random graphs. 
a)~Degree distribution for random triangulations of $8\times8$ (\protect\includegraphics{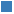}) and $32\times32$ (\protect\includegraphics{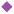}) lattices compared to the average degree distribution of a Newman-Watts (\protect\includegraphics{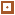}) and a \barabasi (\protect\includegraphics{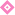}) random graph with $n = 32^2$ vertices.
b)~Specific mean energy, variance of the degree distribution, c)~clustering coefficient and d)~shortest path length for maximal ordered (\protect\includegraphics{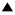}) and random (\protect\includegraphics{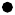}) triangulations, the \erdoes (\protect\includegraphics{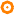}) and the other random graphs in terms of the number of vertices $MN$. For the random triangulations and the Newman-Watts graph the lines were determined by a power-law fit, all other lines are analytical results. The asymptotic behaviour is listed in Tab.\,\ref{tab:random_scaling}.}
\label{fig:results_random}
\end{figure}

In this section we calculate the scaling behaviour of the vertex degree distribution, the mean energy, the average clustering coefficient and the average shortest path length in terms of the system size of random triangulations and compare it with random graphs.
With random triangulations we denote the ensemble with each possible triangulation having the same constant weight and contributes equally in the calculation of expectation values.
The results are especially interesting, because due to the constant ensemble weights (that are in particular independent of the triangulation energies) they do not depend on the actual choice of the energy function.

We calculate these observables in the random ensemble using the Metropolis algorithm \cite{Metropolis_1953} taking into account at least $1000$ different random triangulations per considered system size.
The autocorrelation time of all considered observables is smaller than $10 MN$ Pachner moves, to avoid errors due to the autocorrelation, we execute $1000 MN$ flips before the first and between two successive measurements to calculate the observable averages.

Random triangulations were considered up to a system size of $64\times 64$, which seems to be small if comparing with topological triangulations \cite{Kownacki_2004,Aste_2012}.
For lattice triangulations one has to check the convexity of quadrangles to decide whether a step is permissible, which increases the computation time needed for one Metropolis step and which decreases the step acceptance ratio, because some steps have to be rejected.

In this paper we restrict to quadratic integer lattices for clarity reasons, actually also simulations for non-quadratic lattices were performed.
Except of the average shortest path length (which grows with $|M-N|$) all observables only depend on the actual system size $M\cdot N$ for $M\cdot N \gtrapprox 10^2$ and $M,N \gtrapprox 4$ and not on the two linear sizes.

For comparing the random triangulation averages with the results in the \erdoesnospc, the Newman-Watts and the \barabasi random graphs of the same size that are not known analytically, we use averages over 500 randomly generated instances of the respective random graphs. 
The generation of these random graphs was done using the NetworkX framework \cite{Hagberg_2008}.

\subsection{Vertex degree distribution}\label{subsec:random_degree_distribution}
The vertex degree distribution of the considered random graphs is known analytically: for the \erdoes and Newman-Watts it follows a binomial distribution, for the \barabasi random graph it is a power-law distribution \cite{Dorogovtsev_2000, Krapivsky_2000}.

In figure\,\ref{fig:results_random}a) the degree distribution of random triangulations is displayed for different lattice sizes and compared with the different random graphs.
In contrast to the random graphs the values $k = 0$ and $k = 1$ are not encountered for triangulations since these vertex degrees are forbidden for the graph to form a valid triangulation.
The degree $k = 2$ can only be realised on the boundary of the triangulation, so the probability decreases for increasing lattice size and decreasing importance of the boundary.

The calculations for our random triangulations agree for $k \geq 4$ with the results for the degree distribution of a topological triangulation of a torus in \cite{Kownacki_2004}.
In contrast to our results, where $P(k)$ is peaked at $k = 4$, in \cite{Kownacki_2004} a monotonically decreasing $P(k)$ with maximal value at $k = 3$ was found.
This is because the vertex degree $k = 3$ is difficult to realise due to the non-general coordinates of the vertices in lattice triangulations (there are many collinear points), so even by constructing one cannot realise more than one out of four vertices having degree $k=3$.
In topological triangulations one can realise Apollonian-like networks \cite{Andrade_2005,Zhou_2005,Song_2012}, where every second vertex can have degree $k=3$.

To compare a Newman-Watts random graph we used a ring with each vertex connected to its $L=4$ nearest neighbours and added edges, so for this graph the degrees $k < 4$ are not present.
As a result one can see that the degree distribution of random triangulations is comparable with the \erdoes and Newman-Watts random graph for $k \geq 5$, whereas the \barabasi model shows a scale free power law behaviour $\propto k^{-3}$.

\subsection{Mean energy}
To quantify the disorder of a triangulation we use an energy function which is, up to boundary terms, the variance of the degree distribution multiplied with the lattice size. 
In figure\,\ref{fig:results_random}b) we compare the specific average energy of random triangulations with the degree variance of the random graphs, which can be calculated from the vertex degree distributions.
For the random triangulations a fit to the numerical calculated data yields a convergence of the specific average energy to $4.62 \pm 0.05$, the \erdoes and Newman-Watts random graphs show a similar behaviour (see Tab.\,\ref{tab:random_scaling} for the detailed values).
For the \barabasi random graph the second moment of the degree distribution diverges.

For random topological triangulations the energy per vertex can be calculated analytically to be $E / n = (\langle k \rangle - 3)^2 = 9$ for $\langle k \rangle = 6$ in topological triangulations \cite{Aste_2012}, which agrees with our finding that $E / MN$ converges for $MN \rightarrow \infty$. 
The absolute value of the specific energy in lattice triangulations is smaller, because the degree distribution has a peak located nearer at the mean value as discussed in Sec.\,\ref{subsec:random_degree_distribution}.

\subsection{Clustering coefficient}
The clustering coefficient $C_i = 2K / k_i (k_i - 1)$ of a vertex $i$ with degree $k_i$ is the ratio of the number of connections $K$ between the $k_i$ neighbours and number of possible connections.
For a non-boundary vertex of a triangulation holds $K = k_i$ and $C_i = 2 / (k_i - 1)$, for a boundary vertex $K = k_i - 1$ and $C_i = 2 / k_i$. 
The clustering coefficient $C$ of the whole graph is the average of the vertex clustering coefficients.

The clustering coefficient for the \erdoes random graph equals the edge connection probability $p$ ($\approx 6/MN$ to ensure equal edge number), for the \barabasi model it can be calculated using a mean-field approach \cite{Fronczak_2003}.
In both cases the clustering coefficient vanishes for increasing lattice sizes for our choice of parameters.

The numerical results for the random triangulations and the Newman-Watts graph can be seen in figure\,\ref{fig:results_random}c), both are converging to a constant value for increasing lattice sizes.
For random triangulations the limit is $0.4859 \pm 0.0005$, which is higher than for all considered random graphs (see Tab.\,\ref{tab:random_scaling} for the detailed values).
Intuitively this means that almost every second possible edge between neighbours of a common vertex is present in the random triangulation.

For topological triangulations the clustering coefficient is $C \approx 0.6$ \cite{Aste_2012}.
This is higher than for our lattice triangulations, mainly due to the fact that the degree $k = 3$ is much more probable for topological than for random triangulations.

\subsection{Shortest path length}
As for the clustering coefficient the shortest path length is analytically known for both the \erdoes and the \barabasi random graphs \cite{Fronczak_2004} and has to be calculated numerically for the Newman-Watts graph.
For all random graphs the shortest path length shows small-world behaviour and increases approximately with the logarithm of the vertex number.

These values are compared with the shortest path length of random triangulations in figure\,\ref{fig:results_random}d). 
In contrast to the random graphs, for random triangulations there is a power law behaviour of the shortest path length $\propto (MN)^{0.400 \pm 0.001}$.
Also for random topological triangulations a power law scaling of the average shortest path length can be found \cite{Aste_2012}.

\begin{table*}
  \caption{\label{tab:random_scaling}
    Scaling and functional dependence obtained by a least-square fit of the mean energy (or the degree distribution variance), the clustering coefficient and the shortest path length with the system size for ground state (maximal ordered) and random triangulations as well as for comparable size \erdoesnospc, Newman-Watts and \barabasi graphs. If there are no error bars, the results are taken from analytical calculations.}
  \begin{tabular}{lll}
    quantity & graph & scaling behaviour \\
    \br
    \multirow{5}{*}{$\frac{\langle E \rangle}{MN}$, $\mathrm{Var}(k)$} & ground state triang. & 0 \\
    & random triangs. & $(4.62 \pm 0.02) - (10.80 \pm 0.043)\cdot (MN)^{-0.47 \pm 0.02}$ \\
    & \erdoes & $6 - 4\cdot (MN)^{0.5}$ \\
    & Newman-Watts & $2 - 4\cdot (MN)^{0.5}$ \\
    & \barabasi & $\rightarrow \infty$ \\
    \mr
    \multirow{5}{*}{$\langle C \rangle$} & ground state triang. & $0.4$ \\
    & random triangs. & $(0.4849 \pm 0.0003) - (0.78 \pm 0.02)\cdot (MN)^{-0.73 \pm 0.01}$ \\
    & \erdoes & $6 \cdot (MN)^{-1}$ \\
    & Newman-Watts & $(0.241 \pm 0.001) - (1.18 \pm 0.02)\cdot (MN)^{-0.557 \pm 0.007}$ \\
    & \barabasi & $0.678 \cdot (MN)^{-1}\log(MN)^2$ \\
    \mr
    \multirow{5}{*}{$\langle \ell \rangle$} & ground state triang. & $(0.5677 \pm 0.0002)\cdot (MN)^{0.49978 \pm 0.00005}$ \\
    & random triangs. & $(0.673 \pm 0.006)\cdot (MN)^{0.409 \pm 0.001}$ \\
    & \erdoes & $\log(MN)/6$ \\
    & Newman-Watts & $(0.063 \pm 0.001) \cdot \log\left[ (0.595 \pm 0.002)MN\right]$ \\
    & \barabasi & $\log(MN) / \log(\log(MN))$ \\
    \br
  \end{tabular}
\end{table*}

\section{Canonical triangulations}

\begin{figure}
  \begin{indented}
  \item[]{
      \includegraphics{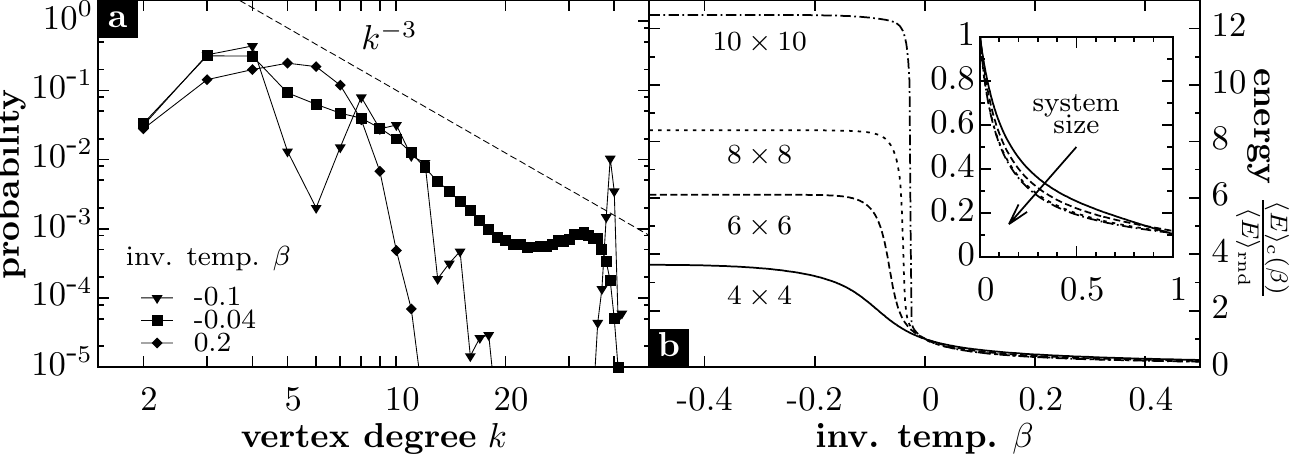}
      \includegraphics{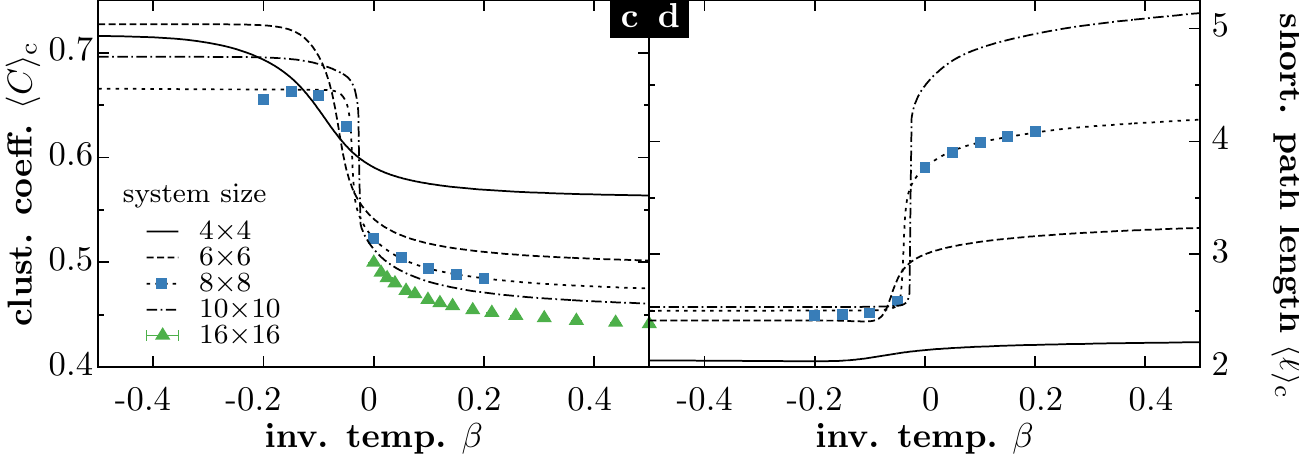}
    }
  \end{indented}
\caption{Temperature dependence of graph properties of lattice triangulations.
  a)~Canonical averaged degree distribution of $8\times 8$ triangulations for different inverse temperatures $\beta$. The dashed line is a power law $\propto k^{-3}$. b)~Mean energy in units of the energy of a random triangulation (with inset for positive temperatures), c)~clustering coefficient and d)~shortest path length in terms of the inverse temperature $\beta$ for triangulations of different lattice sizes. The degree distribution and the lines in b)-d) were calculated using a Wang-Landau simulation, the data points in c)-d) are taken from a parallel tempering simulation.}
\label{fig:results_canonical}
\end{figure}

In this section we consider canonical averages of the degree distribution, the mean energy, the clustering coefficient and the shortest path length of lattice triangulations in terms of the inverse temperature with respect to the energy defined in (\ref{eq:energy}).
These calculations extend the results obtained in the previous section which correspond to the special case of vanishing inverse temperature $\beta = 0$.
The numerical simulations are not restricted to the usually considered case $\beta \geq 0$, but can be extended to the case of negative inverse temperatures $\beta < 0$.
These negative temperatures can also be interpreted as positive inverse temperatures with a negative coupling constant included in the definition (\ref{eq:energy}) of the energy, which makes the triangulations with the most disorder the ground state of the energy function.

It is difficult to use Metropolis Monte-Carlo simulations \cite{Metropolis_1953} to perform canonical averages for triangulations, especially if one wants to explore the regime of negative temperatures/coupling.
There exist triangulations that are local minima in the energy landscape \cite{Knauf_2013}, and due to the small Metropolis acceptance probabilities ($\propto \exp(-\beta \Delta E)$, where $\Delta E$ is the energy difference induced by the flip) the algorithm gets stuck in one of these local minima.
Other states outside of the local minimum that contribute to the ensemble average are reached only after many steps or not at all, so the autocorrelation times and therewith the simulation errors become large, or the possibility arises that the system is not even computationally ergodic anymore.
This problem was also treated analytically in \cite{Caputo_2013}, where the mixing time (which is related to the autocorrelation time) of Glauber dynamics on lattice triangulations is shown to scale exponentially with the system size for a small enough $\beta < 0$.

Using a parallel tempering approach \cite{Earl_2005}, which is basically the parallel calculation of multiple Metropolis simulations at different inverse temperatures with the possibility of interchanging the inverse temperatures, can help to overcome the problem of these local minima in the energy landscape.
But parallel tempering is known to fail in situations with a large free energy barrier, e.g. in first order phase transitions. 

Flat histogram algorithms as the Wang-Landau algorithm \cite{Wang_2001,Wang_2001b} are used to calculate the density of states (DOS) $g(E)$ (which is the normalised number of states with energy $E$) of systems and can help to overcome both the local minima and the free energy barrier problem.
This algorithm samples the single states according to their inverse DOS $g(E)^{-1}$ based on an initial estimation of $g(E)$ and simultaneously improves the estimation until it converges to the actual DOS of the system.
The two main advantages of this algorithm is that with the knowledge of the DOS the observables can be calculated for all inverse temperatures $\beta$ based on only one simulation, whereas for the Metropolis algorithm a new simulation for each $\beta$ has to be used.
The second one is that the algorithm does not get stuck anymore in local minima and the problems of large autocorrelation times described before and in \cite{Caputo_2013} do not occur.
Flat histogram sampling techniques were already applied to two-dimensional lattice triangulations \cite{Knauf_2013}, but are limited to relatively small system sizes.

\begin{figure}
  \begin{indented}
  \item[]{
      \includegraphics{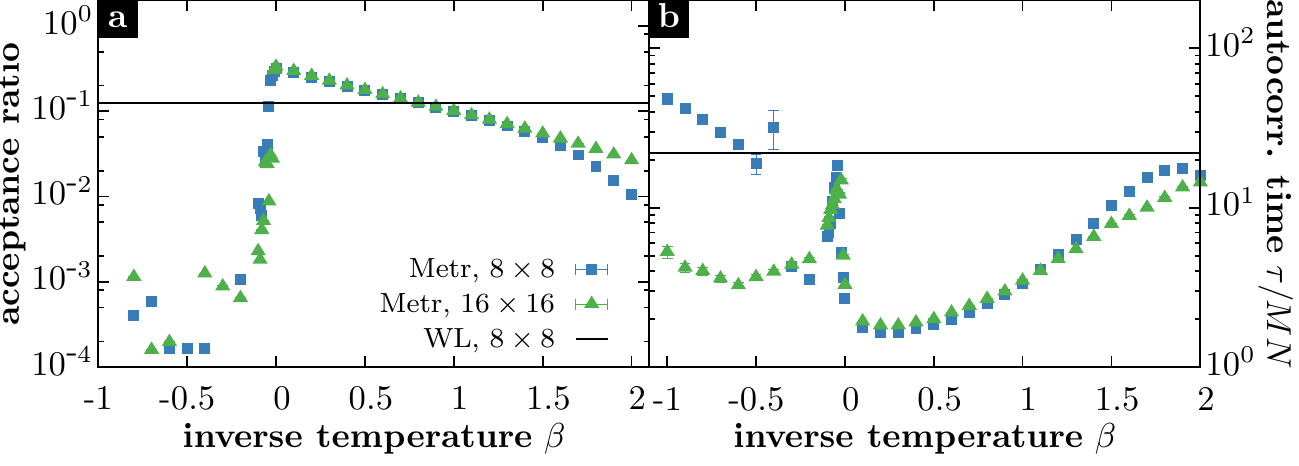}
    }
  \end{indented}
\caption{\label{fig:acceptance_ratio_autocorrelation}Acceptance ratio of a Monte Carlo step (a) and autocorrelation time of the energy observable (b) in terms of the inverse temperature for the Metropolis algorithm for different lattice sizes (points). The solid line is the temperature independent value obtained by a multicanonical simulation. Note that for the simulations based on the density of states the acceptance ratio is lower and the autocorrelation time is higher than in the Metropolis case for certain temperatures ranges, but the Wang-Landau algorithm has on the one hand the advantage that only one simulation has to be performed for the whole temperature range, on the other hand it makes negative temperatures accessible at all.}
\end{figure}

In figure \ref{fig:acceptance_ratio_autocorrelation} one can see a comparison of Metropolis sampling and a sampling based on the DOS calculated with Wang-Landau in terms of the acceptance ratios and the autocorrelation time of the energy observable.
For negative temperatures the Metropolis algorithm basically gets stuck in high-energy states, which leads to low acceptance ratios and long autocorrelation times.
Using the Wang-Landau algorithm this problem does not occur, because steps are weighted according to their entropy difference, and not their energy difference.

The DOS of two-dimensional lattice triangulations can only be calculated up to $11\times 11$ triangulations for all energies, introducing a energy cutoff makes it possible to calculate the density of states up to $25 \times 25$ triangulations that can be used only for positive inverse temperatures $\beta > 0$.
For bigger lattice sizes the Wang-Landau algorithm does not converge anymore in a reasonable amount of time, since on the one hand the entropy between neighbouring states differs by orders of magnitude, on the other hand the high energy states are only connected among themselves by pathes involving low energy states, which makes flat histogram sampling difficult (see \cite{Knauf_2013} for a detailed discussion).

\subsection{Degree distribution}
The canonical averaged distribution $P(k)$ of the vertex degrees of $8 \times 8$ lattice triangulations is plotted for different temperatures in figure\,\ref{fig:results_canonical}a).
For the positive inverse temperature $\beta = 0.2$ one finds a similar behaviour as for the random triangulations $\beta = 0$ with the maximum of the degree distribution shifted towards degrees $k = 5$ and $k = 6$, because this degrees are preferred at low temperatures due to the energy function (\ref{eq:energy}).
For negative inverse temperatures $\beta = -0.1$ there is an additional peak at vertex degree $k \approx 40$, because for this inverse temperature the triangulation with maximal energy contributes most to the ensemble average.
An interesting behaviour can be found for the inverse temperature $\beta = -0.04$ (here the probability distribution of the triangulation energies has two peaks, which is a hint for a phase transition).
For this inverse temperature the degree distribution behaves similar to a power law $k^{-3}$ for vertex degrees in the range $5 < k < 25$, which matches the degree distribution of the \barabasi model qualitatively.
One has to admit that due to the small system size the range of vertex degrees found spans less than one magnitude and that there are deviations from a pure power law behaviour even in this range.
Nevertheless the qualitative change of the degree distribution compared to random triangulations is remarkable.
Additionally we suspect that for bigger system sizes there is a clearer power law behaviour for larger ranges of degrees.

Similar results for the canonical averaged triangulation degree distribution can be found in \cite{Kownacki_2004} for positive temperatures.

\subsection{Mean energy, clustering coefficient and shortest path length}
To compare the temperature dependence of the mean energy for different lattice sizes, we consider the expectation value of the mean energy $\epsilon = \langle E \rangle_{\mathrm{c}} (\beta) / \langle E \rangle_{\mathrm{rnd}}$ in terms of the energy of a random triangulation.
The results of the numerical calculations for the relative mean energy, the clustering coefficient and the shortest path length can be found in figure\,\ref{fig:results_canonical}b) - d).

The temperature dependence of all considered observables shows the characteristic behaviour of a first-order phase transition, with quasi-critical inverse temperature $\beta_c \rightarrow 0$ for increasing system size, but this is not a real phase transition.
For the maximal energy of a quadratic $N\times N$ lattice triangulation there is a lower bound that scales with $N^4$ \cite{Knauf_2013}, so that the maximal specific energy scales at least with $N^2$. 
For $\beta < 0$ (which is equivalent to using negative coupling) these maximal energy states are the ground states, which are then not bounded by below for the limit of infinite system size.
For negative temperatures only finite systems can be considered and and the thermodynamic limit of infinite system size cannot be obtained, so the behaviour found at $\beta \rightarrow 0$ is then no actual phase transition in sense of statistical physics.

For negative inverse temperatures one finds a high clustering coefficient (approximately 0.7 independent of the system size), and small average path length between 2 and 3 independent of the system size.
The latter can be understood in terms of the graph theoretical diameter of the highest energy triangulations, which can be shown to be 4 independent of the lattice size.
So for a small enough negative inverse temperature one can find a small-world behaviour for the triangulation graphs.

\begin{figure}
  \begin{indented}
  \item[]{
      \includegraphics{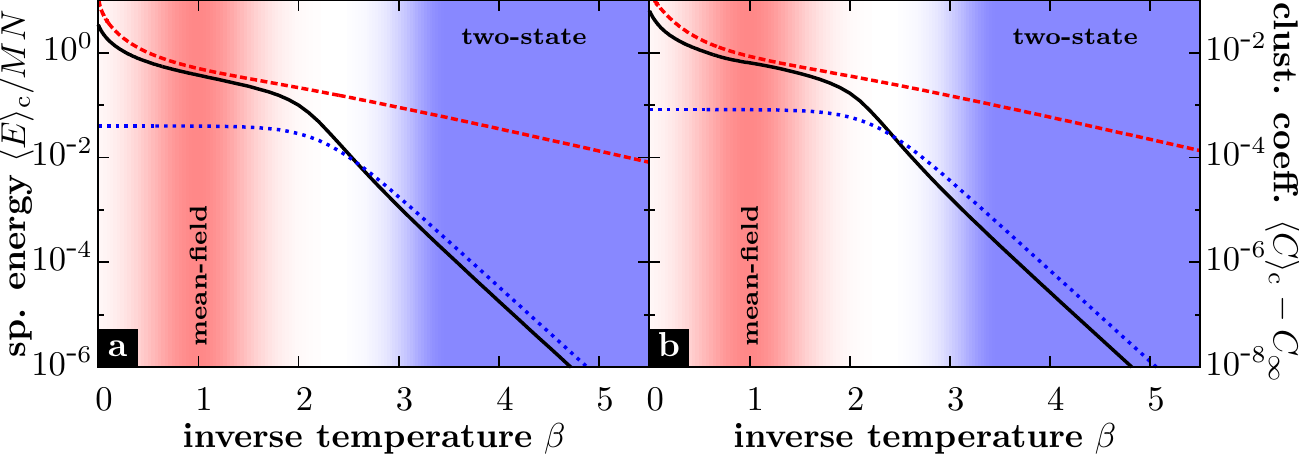}
    }
  \end{indented}
\caption{\label{fig:results_canonical_analytical}Analytical approximations for the specific energy (a) and the clustering coefficient (b) in two different regimes compared with the actual data obtained numerically. The mean-field approach (red, dashed line), which is valid for an intermediate regime of positive inverse temperatures, neglects the entropic properties of the triangulations, the two-state approximation (blue, dotted line), which is valid for $\beta \rightarrow \infty$, just considers the ground state and the first excited state of the lattice triangulations.}
\end{figure}

For positive temperatures the average energy per vertex and the clustering coefficient can be approximated analytically.
The first possibility is to use the mean-field approach suggested in \cite{Aste_2012}, where the topological and entropic properties of the triangulations are neglected.
This results in 
\begin{equation*}
  \frac{\langle E \rangle}{MN} \approx \frac{2}{2 + e^\beta} \quad \langle C \rangle \approx \frac{ 5e^{-\beta}/6 + 2/5}{2 e^{-\beta} + 1}
\end{equation*}
The second possibility is to use the known degeneracy $\Omega(E = 4)$ and $\Omega(E = 0)$ of the ground state and the first excited state \cite{Knauf_2013} to calculate the expectation values neglecting all other states, resulting in
\begin{equation*}
  \begin{split}
    \frac{\langle E \rangle}{MN} &\approx \frac{ \Omega(E = 4) }{MN} \frac{4}{e^{4\beta} + \Omega(E = 4)} \approx \frac{4 MN}{2e^{4\beta} + M^2N^2} \\
    \langle C \rangle & \approx C_{\mathrm{gs}}\frac{1 + (15 MN)^{-1} \Omega(E = 4)e^{-4\beta}}{1 + \Omega(E = 4)e^{-4\beta}}
  \end{split}
\end{equation*}
Both approximations are compared with the numerical results in figure\,\ref{fig:results_canonical_analytical}.
For both the specific energy and the clustering coefficient the mean field approximation is correct for $0 < \beta < 1.5$, the two-niveau low-temperature approximations is correct for $\beta > 3$.

\section{Conclusion}

In this paper we proposed a new model for real-world graphs or networks by interpreting unimodular triangulations of two-dimensional integer lattice as graphs.
Considering averages of random triangulations we found that they show a higher clustering coefficient than common random graph models, but also a power law growing shortest path length in terms of the system size, where one would expect a logarithmic scaling for small world behaviour. 
The degree distribution behaves similar as in the \erdoes and Newman-Watts random graph.

Introducing an energy function that measures the order and disorder of a triangulation, canonical averages of graph observables in triangulations can be calculated using the Wang-Landau algorithm.
This Monte Carlo method made it possible for the first time to calculate expectation values for the whole temperature range, albeit only for small system sizes.
Using the Wang-Landau algorithm it is possible to access equilibrium properties of both high energy triangulations, which are poorly connected with other similar triangulations, and low energy triangulations, where dynamically a glass-like phase was found in the literature before, both phenomena making it difficult to use the Metropolis algorithm.
Considering the temperature dependence of the clustering coefficient, the average shortest path length and the degree distribution we found small-world behaviour for negative temperatures $\beta < \beta_c$ below a negative, quasi-critical temperature and hints for scale-free behaviour for temperatures near the quasi-critical temperature.
All considered observables show a cross-over behaviour going from negative temperatures, where the partition function dominated by disordered triangulations, to positive temperatures, where it is dominated by ordered triangulations.

In contrast to the topological the lattice or in general embedded triangulations are useful if one deals with networks and graphs where the actual coordinates of the vertices become important.
Despite the introduction of vertex coordinates which induce non-executable flips, our results show qualitatively agreement with the graph properties of topological triangulations of surfaces with suitable genus that were estimated analytically and calculated numerically in the literature.

An extension to the grandcanonical ensemble, where the number of vertices is not fixed any more, is possible by considering also insertion and removal Pachner moves. Additionally one can apply the methods described in this paper also to triangulations of arbitrary point sets in two and more dimensions.

\section*{Acknowledgements}
The authors thank J. F. Knauf, R. Jonsson and A. Kraynik for fruitful discussions. This work is supported by EFI Quantum Geometry and the Elite Network of Bavaria.

\bibliographystyle{iopart-num}  
\bibliography{literature_laplacian}

\end{document}